\documentclass[a4paper,twocolumn,11pt]{quantumarticle}
\pdfoutput=1
\usepackage[numbers,sort&compress]{natbib}
\usepackage[utf8]{inputenc}
\usepackage[english]{babel}
\usepackage[T1]{fontenc}
\usepackage{amsmath,amsfonts}
\usepackage{hyperref}
\usepackage{booktabs}
\usepackage{tikz}
\usepackage{lipsum}
\usepackage{subcaption}

\usepackage[linesnumbered,ruled,vlined]{algorithm2e}

\usepackage{longtable}
\usepackage{braket} 
\newtheorem{prop}{Proposition}
\newtheorem{definition}{Definition}

\begin{document}

\title{Orbit classification and analysis of qutrit graph states under local complementation and local scaling}

\author{Konstantinos-Rafail Revis}
\email{krevis@mail.uni-paderborn.de}
\orcid{0000-0002-5645-5448}
\affiliation{Department of Computer Science, Paderborn University, Warburger Str. 100, 33098, Paderborn, Germany}
\affiliation{Institute for Photonic Quantum Systems (PhoQS), Paderborn University, Warburger Str. 100, 33098 Paderborn, Germany}
\author{Hrachya Zakaryan}
\email{hzak@mail.uni-paderborn.de}
\orcid{0009-0009-2862-2243}
\affiliation{Department of Computer Science, Paderborn University, Warburger Str. 100, 33098, Paderborn, Germany}
\affiliation{Institute for Photonic Quantum Systems (PhoQS), Paderborn University, Warburger Str. 100, 33098 Paderborn, Germany}
\author{Zahra Raissi}
\email{zahra.raissi@uni-paderborn.de}
\orcid{0000-0002-9168-8212}
\affiliation{Department of Computer Science, Paderborn University, Warburger Str. 100, 33098, Paderborn, Germany}
\affiliation{Institute for Photonic Quantum Systems (PhoQS), Paderborn University, Warburger Str. 100, 33098 Paderborn, Germany}

\begin{abstract} 

Graph states and their entanglement properties are pivotal for the development of quantum computing and technologies. For qubits, local complementation, a graphical rule that connects all the equivalent states under Local Clifford (LC) operations, was used for the complete characterization of all the LC equivalence classes up until 12 particles, assisting applications in quantum error correction and state preparation protocols optimization. This concept has been extended for qudits. In this work, we provide a complete characterization of the entanglement classes up until $7$ qutrits, mapping each class into an orbit. The graph-theoretic properties of the orbits are studied, illuminating the rich structure they have. Clear connections between the connectivity of the orbits and the entanglement properties are observed. The correlations between the graph-theoretic properties and the Schmidt measure provide useful insights regarding qudit state preparation and fault-tolerance. This work is accompanied by a repository consisting of the code to extract the orbits and perform the presented statistical analysis. The strong interplay between quantum theory and graph theory is well-known and extensively studied for qubits. Our work provides practical tools and results to assist this endeavor for the qudit case.

\end{abstract}
\maketitle
\section{Introduction}\label{sec:intro}

Graph states provide a natural framework to study entanglement, which lies at the core of quantum computation, quantum communication, and quantum error correction across the most prominent physical realizations \cite{raus_intro,Veldhorst2017,trapped_ion,intro_photons,intro_multiplexed, Barends2014}. 
Both qubit and qudit graph states also have applications in quantum secret sharing and measurement-based quantum computing \cite{secret_sharing, qudit_secret_sharing, raus_intro}.

In the groundbreaking works \cite{lc_paper, algo_van_de_nest}, it was shown for the qubit case that every Local Clifford (LC) operation on a graph state can be represented as a series of graphical rules known as local complementation. The quantum operations corresponding to local complementation were described in \cite{Eisert}, along with a classification of all entanglement classes (LC orbits) for qubit graph states with up to 7 qubits. This catalog was extended for $n\leq 8$ in \cite{reps_8} and ultimately for $n\leq 12$ in \cite{qubits_reps12}. In \cite{Adcock2020mappinggraphstate}, the idea to represent the entanglement classes as orbits and study the corresponding graph properties was introduced, which turned out to be an interesting interpretation, finding various theoretical and experimental applications.

Recent works have generalized local complementation to include non-Clifford transformations \cite{adamLU, rlocal,claudetLU}. These generalizations enabled the discovery of counterexamples to the LU-LC conjecture \cite{ji2008lulcconjecturefalse} up to 19 qubits using a pseudo-polynomial algorithm.
Local complementation has been additionally extended to include qudit systems of prime local dimension \cite{beigiLC, quditorbitalgorithm,quditLC1}, where the graph states are described as multi-graphs or weighted graphs. Moreover, an algorithm to check if two qudit graph states are equivalent under some sequence of local complementations was given in \cite{beigialgorithm}. In the qutrit case $d=3$, \cite{qutrits_reps12} provided representative graphs for the LC orbits up to $n=12$, along with an efficient algorithm for their computation.

In the monumental work by Schlingemann and Werner \cite{Schlingemann_2001}, the construction scheme of quantum error correction codes based on a graph and a finite abelian group was established. Many recent works \cite{Lobl2024losstolerant,PRXQuantum_6_010305,wu2024,khesin2024equivalenceclassesquantumerrorcorrecting} emphasize the need for lookup tables of LC orbits, both for efficient state preparation and fault-tolerant implementations.
Additionally, the study of fault tolerance of concatenated graph codes was highlighted in \cite{PRXQuantum_6_010305} and \cite{ft_graph_codes}. One has to underline that the aforementioned works are based on qubits. Our work provides a key ingredient for the extension of these works for qutrits and, consequently, for every prime local dimension.

While the literature on fault-tolerance and quantum error correction for qudits is less developed than for qubits, it has been shown that increasing the local dimension can enhance code performance \cite{PhysRevLett_113_230501}.
Additionally, leveraging qudit systems boosts the scalability of the systems and the application of logical operators \cite{PhysRevX_2_041021}. Technical details on the transition already exist in the literature \cite{imec_qudit_qec}, as well as the smallest stabilizer code to correct any single qudit error has been presented. Our work provides a necessary step to further optimize error mitigation schemes or assist in the extension of the already existing ones to the qudit case.

Another context in which being equipped with a lookup table is useful, is graph state preparation. For qubits, several works have used orbits to minimize circuit depth and gate count \cite{opt_grap_state_prep,perdrix_state_prep, takou2024, kumabe2025, PhysRevA_110_052605}. Additionally, further optimization approaches utilize fusion-based schemes, but local complementation remains a key ingredient to reduce the number of fusions and consequently the required resources to prepare the state \cite{PhysRevA_111_052604_fusion_lc}. Therefore, extending this lookup table to any prime dimension provides a crucial step to the qudit extension of optimization of state preparation protocols.

As shown in \cite{beigisize} both the number of qudit graph state orbits and the size of the orbits on average grows exponentially, more specifically as $O(d^{n^2/2})$ for the number of orbits and as $O(d^{2n})$ for the orbit size, where $d$ is the local dimension. For qubits finding the orbits is a \#P-Complete problem \cite{dahlberg2019complexityvertexminorproblem, Dahlberg2018}.  Therefore, due to the first argument and the fact that the core structure of the problem is similar for both qubits and qudits, fully classifying the orbits is computationally difficult from both time and space perspectives. For qubits, full orbit characterization has been done up to 12 qubits, and a detailed statistical analysis up to 9.

In this work, we extend the orbit classification to qutrit ($d=3$) graph states. However, even with the use of parallelization to speed up our algorithm, we reach up to $n=7$, due to the exponential nature of the problem from both time and space standpoints. We believe $n=9$ is reachable using a high-performance computing cluster. Additionally, we provide the statistical analysis of all of these orbits. The close connection between the orbit connectivity and the corresponding entanglement properties is demonstrated. We find that the correlations between orbit properties closely follow those of the qubit case. The code for all the computations is available at \cite{github}.

The paper is structured as follows. In section \ref{sec:qudit-gs} we provide the definitions of qudit graph states, and local complementation. In section \ref{sec:methods} we outline the methods used to obtain the data, while in section \ref{sec:results} we provide the results and discuss the physical implications. We conclude and discuss the results in section \ref{sec:discussion}.

\section{Qudit Graph States}\label{sec:qudit-gs}
Qudit graph states over a prime dimension $d$ are naturally described using weighted graphs or multigraphs (without self-loops), where edge weights take values in the finite field $\mathbb{F}_d$. These descriptions are equivalent and will be used interchangeably throughout.

Consider a weighted graph $G(V,E)$ with weights from $\mathbb{F}_d$ for a prime $d$, where $V$ is the set of vertices and $E$ is the set of edges with their associated weights. Additionally, the graph can be described by the adjacency matrix $\Gamma$,
\begin{align}
    \Gamma_{uv}=\begin{cases}
        r & \text{if } \{u,v,r\} \in E\\
        0 & \text{otherwise}.
    \end{cases}
\end{align}
We define the graph state associated with the graph $G$ as,
\begin{align}
    \ket{G}=\displaystyle\prod_{\{u,v\} \in E} CZ_{uv}^{\Gamma_{uv}}\ket{+}^{\otimes n},
\end{align}
where $\ket{+}=H\ket{0}$ and $CZ$ is the qudit controlled-Z operation,
\begin{align}
    CZ\ket{kj}=\omega^{kj}\ket{kj},
\end{align}
with $\omega=\exp(\frac{2\pi i}{d})$.

\subsection{Local Scaling and Complementation}
It is well known that for qubit systems, graph states can be separated into Local Clifford equivalent classes \cite{lc_paper,Eisert}. We say that two qubit graph states are LC equivalent if their graphs can be transformed to each other using a sequence of local complementations, where local complementation on a vertex $k$ is given by,
\begin{align}
    \Gamma_{ij} = \Gamma_{ij}+\Gamma_{kj}\Gamma_{ki} \text{ (mod } 2),\ \forall i,j,
\end{align}
meaning we complement the neighborhood of the vertex $k$.\\
It has been shown that this process can be extended to prime dimensional qudit graph states \cite{quditLC1,beigiLC}, allowing us to graphically describe generalized Clifford transformations on graph states. For qudit systems we consider two types of graphical operations, local scaling and local complementation. Note that from here on all the computations are done over $\mathbb{F}_d$.
\begin{prop} \cite{quditLC1} (Local Scaling)
    For any $\gamma \in \mathbb{F}_d$, the $\gamma$-local scaling about a vertex $w$ in a graph $G$ is given by,
    \begin{align}
        (\Gamma\overset{\gamma}{\circ} w)_{uv}=\begin{cases}
            \gamma \Gamma_{uv} & \text{if } u=w \text{ or }  v=w\\
            \Gamma_{uv} & \text{otherwise}.
        \end{cases}
    \end{align}
    The graph state transforms under $\gamma$-local scaling as,
\begin{align}
    \ket{G \overset{\gamma}{\circ} w} = M_w(\gamma^{-1})\ket{G},
\end{align}
where $M(\gamma)\ket{k}=\ket{\gamma k}$.
\end{prop}

\begin{figure}
    \centering
    \includegraphics[width=0.9\linewidth]{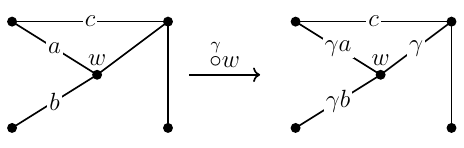}
    \caption{Local scaling with scaling factor $\gamma$ on the vertex $w$.}
    \label{fig:scaling}
\end{figure}

\begin{figure}
    \centering
    \includegraphics[width=0.9\linewidth]{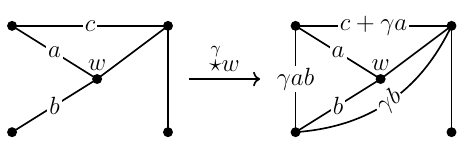}
    \caption{Local complementation with scaling factor $\gamma$ on the vertex $w$.}
    \label{fig:complementation}
\end{figure}

 \begin{prop} \cite{quditLC1} (Local Complementation)
     For any $\gamma \in \mathbb{F}_d$, the $\gamma$-local complementation about a vertex $w$ in a graph $G$ is given by,
     \begin{align}
         (\Gamma \overset{\gamma}{\star} w)_{uv}=\begin{cases}
             \Gamma_{uv}+\gamma \Gamma_{uw}\Gamma_{vw} & \text{if } u\neq v\\
             \Gamma_{uv} & \text{otherwise}.
         \end{cases}
     \end{align}
     The graph state transforms under $\gamma$-local complementation as,
     \begin{align}
         \ket{G \overset{\gamma}{\star} w}=\Tilde{P}^\gamma_w\displaystyle\prod_{v\in V}P_v^{-\gamma \Gamma^2_{vw}}\ket{G},
     \end{align}
     where,
     \begin{align}
         P\ket{x}=\omega^{\frac{x(x-1)}{2}}\ket{x},
     \end{align}
     and
     \begin{align}
         \Tilde{P}=H^\dagger PH.
     \end{align}
     
 \end{prop}
Examples of local scaling and local complementation are given in Fig.~\ref{fig:scaling} and Fig.~\ref{fig:complementation} respectively.

Note that for qudits the generalized Clifford group is defined as the normalizer of the generalized Pauli group,
\begin{align}
    \mathcal{G}_n=\{\omega^c X(\bold{a})Z(\bold{b}): \bold{a},\bold{b}\in \mathbb{F}_d^n, c\in\mathbb{F}_d\},
\end{align}
where $X(\bold{a})=X^{a_1}_1\otimes X^{a_2}_2\otimes \cdots \otimes X^{a_n}_n$ and $Z(\bold{a})=Z^{a_1}_1\otimes Z^{a_2}_2\otimes \cdots \otimes Z^{a_n}_n$ for $\bold{a}=\{a_1,a_2,...,a_n\}$. Based on this, it is simple to show that, $M(\gamma)$, $P$ and $\Tilde{P}$ are generalized local Clifford operations.

\section{Methods}\label{sec:methods}
\subsection{Orbit Extraction}

Obtaining the orbits is done in three steps. First, we create all possible non-isomorphic weighted graphs. Next we create an atlas of graphs, where the vertices are all the non-isomorphic weighted graphs and the edges all the possible local scalings and local complementations. Lastly, we separate the disjoint graphs to obtain the orbit graphs.

In more detail we do the following. Note that from here on we will deal with qutrits specifically. We generate all the non-isomorphic simple graphs of up to 7 vertices using \texttt{nauty}'s \texttt{geng} \cite{nauty}. Next we need to add the weights. \texttt{geng}'s output is in Graph6 format which doesn't support weighted graphs, hence we instead use our own encoding. We consider the values in the adjacency matrix with their bit representations and treat the vectorized upper triangle as a $b\times n(n-1)/2$ size binary number. Here $b$ is the number of bits required to represent each weight, which for qutrits is $2$, as the possible values are $0$,$1$ and $2$. Hence, our encoding gives us an integer for each graph. For each simple non-isomorphic graph we generate all possible weight combinations, and choose a canonical form by generating all the possible permutations for each weight combination and taking the smallest encoding. We parallelize this process using the Python \texttt{multiprocessing} library as we can run the process for each simple graph independently. Next, we create a graph atlas with all the generated graphs as vertices. For each graph we perform a single layer of all possible local scalings and local complementations for all scaling factors and vertices in the graph and we add these as edges to the atlas. Once we have done this for all the graphs we will have an orbit atlas for all the possible local scalings and local complementations. As the creation of each layer is independent, once again, we can parallelize. Lastly, we use the \texttt{networkx}'s \texttt{connected\_components} function to separate the disjoint orbits in the orbit atlas, which is done through a breadth-first search and has a complexity of $O(V+E)$.\\
As the number of non-isomorphic graphs scales exponentially \cite{beigisize}, so do the time and space complexities of our algorithm. Therefore, we find the orbits for $d=3$ and $n\leq 7$. For the future, we believe it should be possible to reach $n=9$ for $d=3$ using high-performance computing clusters.
\subsection{Graph state orbits, observables and statistics}

\subsubsection{Entanglement classes as graphs}
By applying a series of local complementation and local scaling operations, one can obtain every graph state belonging to the same LC entanglement class. Determining which states belong to the same class is not the only information we can extract. As indicated by \cite{Adcock2020mappinggraphstate}, the application of local complementation and local scaling operations creates an orbit. Therefore, it can be represented as a graph, which we are going to call an orbit graph (OG), with the vertices of the graph being the graph states and the edges between two vertices indicating that we can transform one graph state to another by applying a local scaling or a local complementation to one of the vertices of the first. In contrast with the qubit case, it is possible to represent the two different types of local operations and the corresponding weights as edge weights of the OG. The reason we do not choose to proceed in this direction is that the OG will then need to become a weighted graph, making the computation of some of its characteristic properties rather complex, with no significant improvement in the information about the orbit and the physical properties of the entanglement class. For this reason, two vertices will be connected with a single edge independently of the weight or the type of transformation. Furthermore, we will only check non-isomorphic graphs, meaning we won't consider the labeling of the graphs, due to computational limitations. Additionally, the OG is an undirected graph since the local operations are self-inverse. One last remark is that a single application of one of the local operations can lead to the same states and thus, OG can have self-loops.

\subsubsection{Observables}\label{subsection}
We will consider several quantities and concepts related to graph theory. These include chromatic number, density, degree, number of self-loops and diameter. Other than the graph-theoretic quantities we also consider the Schmidt measure entanglement monotone. More detailed description can be found in appendix \ref{app:quant-def}. In general, we are going to call these quantities observables, and they are separated into three categories. The first one is the Schmidt measure, a purely quantum quantity and arguably the most important for us. The second one is graph-theoretic observables of the OG. And the third one is graph-theoretic observables of the graph states, where care must be taken due to edge weights.

Let us start by discussing the chromatic number of OG ($\chi_{OG}$), as well as the minimum chromatic number of any graph state within its orbit ($\chi_i$). For this calculation we make use of the \texttt{networkx} function \texttt{coloring.greedy\_color}. The strategy used was the largest first, and it is described in \cite{graph_coloring}. The number of self-loops ($N_{\mathcal{L}}$) in a graph is calculated by the function \texttt{number\_of\_selfloops}. The calculation of the density of a graph ($\mathcal{D}$) is straightforward since it requires only the number of edges and vertices of the graph as indicated in the corresponding definition in Appendix \ref{app:quant-def}. 

This is the point to depart from the wonderland of the computationally light calculations. The next two observables are related to the distance matrix. Namely the Average Shortest Path Length (ASPL) denoted as $\langle d_{OG} \rangle$ and the diameter of the OG which is $\max\limits_{\forall i,j \in E} \{ d_{OG}^{ij} \}$, frequently denoted as $d_{OG}^{\text{max}}$. The distance matrix was introduced in \cite{distance_matrix_intro}. The ASPL for every OG for $n\leq 6$ and more than half of the OG for $n=7$ is calculated exactly according to the function \texttt{average\_shortest\_path\_length} provided by \texttt{networkx}. For the rest, a heuristic method has been applied. Our approach was instead of calculating every possible shortest path, randomly choose two nodes, calculate their shortest distance path, and then calculate the average. For this, we randomly sampled different subsets and then averaged over them, obtaining the final ASPL for the corresponding OG, giving a rather good approximation compared to the exact one but requiring much less computational time. 

Let us proceed with the calculation of the diameter of the OG. Once again, this task becomes computationally challenging for large graphs. The approach is pretty similar to the ASPL calculation. The idea is to randomly select nodes of the OG and start a breadth-first search. In our case, we chose randomly $1000$ vertices and found the most distant node from them. Once again, for every orbit with $n\leq 6$ and plenty of those with $n=7$ we used the built-in \texttt{networkx} function \texttt{diameter} to precisely calculate their diameter. The reason for that was to benchmark our method by comparing our approximate results with the exact ones. Our results are in perfect agreement with \texttt{networkx}, indicating that we are getting at least a pretty good estimate of the diameter for the larger OGs if not the exact answer.

Another interesting graph-theoretic property is the edge coloring or chromatic index and the vertex degree, which we define in Appendix \ref{app:quant-def}. There is a well-known result in graph theory connecting the maximum degree of a vertex in the graph $G$ ($\text{deg}(G)_\text{max}$) and the chromatic index, the \textit{Vizing's theorem} \cite{vizing_theorem}. According to it, any graph can be colored with at most $\text{deg}(G)_\text{max} +1$ colors. If graphs with multiplicity $\mu$ are assumed, the graph can be colored with at most $\text{deg}(G)_\text{max} + \mu$ colors. In our context, when qudit graph states are described, $\mu = d-1$. Considering that the edge coloring is an NP-complete problem \cite{edge_color_npcom}, determining it for the OG is a computationally costly task. However, the calculation of the maximum degree is not. Therefore, we will focus on the maximum degree of a vertex of the OG, denoted as $\text{deg}(OG)_{\text{max}}$. For every multi-graph in an orbit, the maximum degree is defined as $\max\limits_{\forall g \in \text{orbit}} \{ \deg(g) \}$. Then the minimum value is obtained $\text{min}\left\{\max\limits_{\forall g \in \text{orbit}} \{ \deg(g) \}\right\}$, denoted as $\deg(g)_{\text{min}}$.
It has to be underlined that we were able to calculate the minimum chromatic index of the graph states in a specific entanglement class using the Wolfram Engine, but since obtaining edge coloring for the OG was computationally expensive, in our correlation analysis we keep the $\deg(g)_{\text{min}}$ for consistency. 

\subsubsection{Schmidt Measure Computation}
The Schmidt measure, introduced in \cite{sm-def}, is an entanglement monotone applicable to both qubit and qudit systems. While its connection to graph properties was originally established for qubit graph states, the generalization to prime-dimensional qudits is straightforward \cite{our_col_paper}.
\begin{definition}
    For a state $\ket{\psi}$ the Schmidt measure is defined as,
    \begin{align}
        E_S(\ket{\psi})=\log_d (r),
    \end{align}
where $r$ is the Schmidt rank of the state.
\end{definition}
A more formal definition can be found in Appendix A. In the usual definition $\log_2$ is used, but we use $\log_d$ here to make comparisons between different $d$ values more clear. This does not change the features of the Schmidt measure as it is simply a scaling. Additionally, using $\log_d$ helps with calculating the Schmidt measures based on the colorability of the graph as shown in \cite{our_col_paper}.

For our algorithm we use the following features of the Schmidt measure,
\begin{enumerate}
    \item It is bounded from below by the maximum rank of all possible bipartitions of the adjacency matrix.
    \item As it is an entanglement monotone, it doesn't change under local operations and hence is the same for all graph states in the orbit.
    \item Upon the removal of a vertex the Schmidt measure can at most decrease by 1.
    \item If the graph is two colorable the Schmidt measure is bounded from below by half the rank of the adjacency matrix, and by $\left\lfloor \frac{n}{2} \right\rfloor$ from above.
    \item If the graph is two colorable and the adjacency matrix is full-rank, the Schmidt measure is $\left\lfloor \frac{n}{2} \right\rfloor$.
    \item If the graph is three colorable and the number of vertices with lowest color are $n_1$ and $n_2$, then we can reduce the state to $d^{n_1+n_2}$ terms, meaning that if the lower bound agrees with this value we have an exact Schmidt measure value.
\end{enumerate}
For most orbits an exact value can be found, however for some orbits only an upper and lower bound can be found.

\subsubsection{Correlation Coefficients and special case of our results}\label{subsection:correlation-coef}
Following the thought process presented in \cite{Adcock2020mappinggraphstate}, we would like to examine the correlation of the extracted observables and discuss their physical implications. For this reason, we are going to use the Pearson correlation coefficient. In general, for two random variables $x$ and $y$ the Pearson coefficient $r$ is defined as:
\begin{equation}
    \label{eq;pearson-def}
    r= \frac{\text{cov}(x,y)}{\sigma_x\sigma_y},
\end{equation}
where $\text{cov}(x,y)$ is the covariance of the two variables, while $\sigma_x$ and $\sigma_y$ are the standard deviations of the variables $x$ and $y$. The Pearson correlation coefficient measures the linear relationship between two data sets. In general, $-1\leq r\leq 1$ and if $r=0$ no correlation is implied. In case of $r=1$ an exact linear correlation is implied, while for $r=-1$, an exact linear anticorrelation is validated. A final remark is that if $|r|>0.7$, a strong (anti)correlation is typically assumed. 

Another correlation coefficient used in this work is Kendall's $\tau$. We include it alongside Pearson’s coefficient for several reasons. Firstly, there is no physical reason to believe that the observables we are interested in have only linear correlations. The two observables may have nonlinear monotonic behavior. In this case the Pearson coefficient will not detect any correlation, while the Kendall's $\tau$ will. Secondly, some of the observable data have outliers and knots. Pearson's coefficient is sensitive to them while Kendall's $\tau$ is more robust. Thirdly, since isomorphic graphs are not considered, our data set is smaller, and thus it is possible to violate the assumptions for the Pearson coefficient, leading to disproportionate influence. On the contrary, once again Kendall's $\tau$ is more powerful and robust. Finally, using two correlation coefficients is great to validate our findings and indicate the existence of a nonlinear correlation between the studied observables.

The Kendall's $\tau$ used in this work is defined as \cite{kendall1945ties}:
\begin{equation}
    \tau = (p-q)\left((p+q+t_x)(p+q+t_y)\right)^{-1/2},
\end{equation}
with $p$ the number of concordant pairs, $q$ the number of discordant pairs, $t_x$ the number of ties only in $x$ and $t_y$ the number of ties only in $y$. Additionally, if a tie appears at the same pair for both $x$ and $y$, then it is not considered in either $t_x$ or $t_y$. Finally, $|\tau|>0.7$, is an indicator of strong correlation between the two datasets.

\section{Results}\label{sec:results}

\subsection{Orbit Examples}
\begin{figure*}
    \centering

    \includegraphics[width=\textwidth]{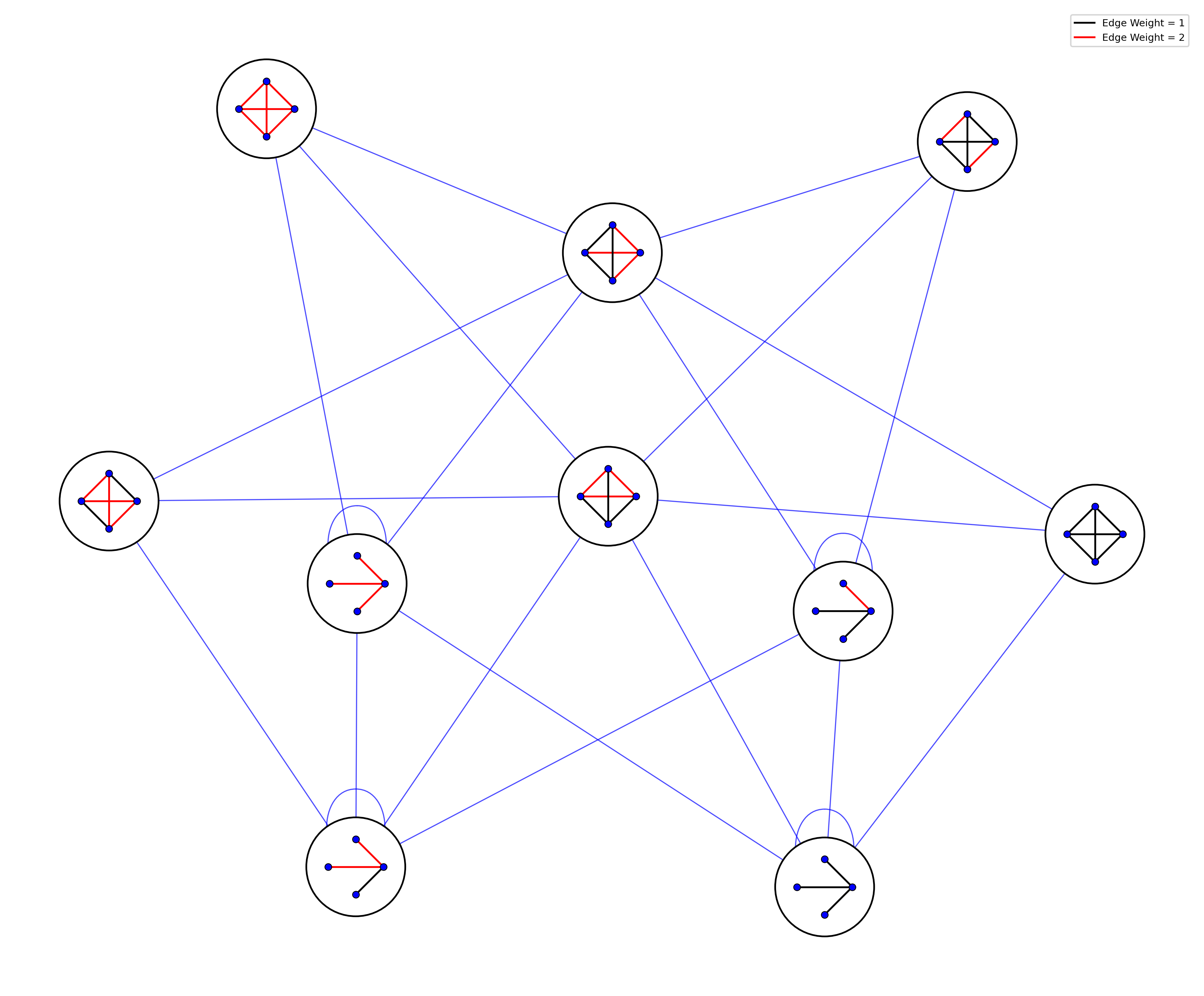}
    \vspace{1em}

    \begin{minipage}[t]{0.49\textwidth}
        \centering
        \includegraphics[width=\columnwidth]{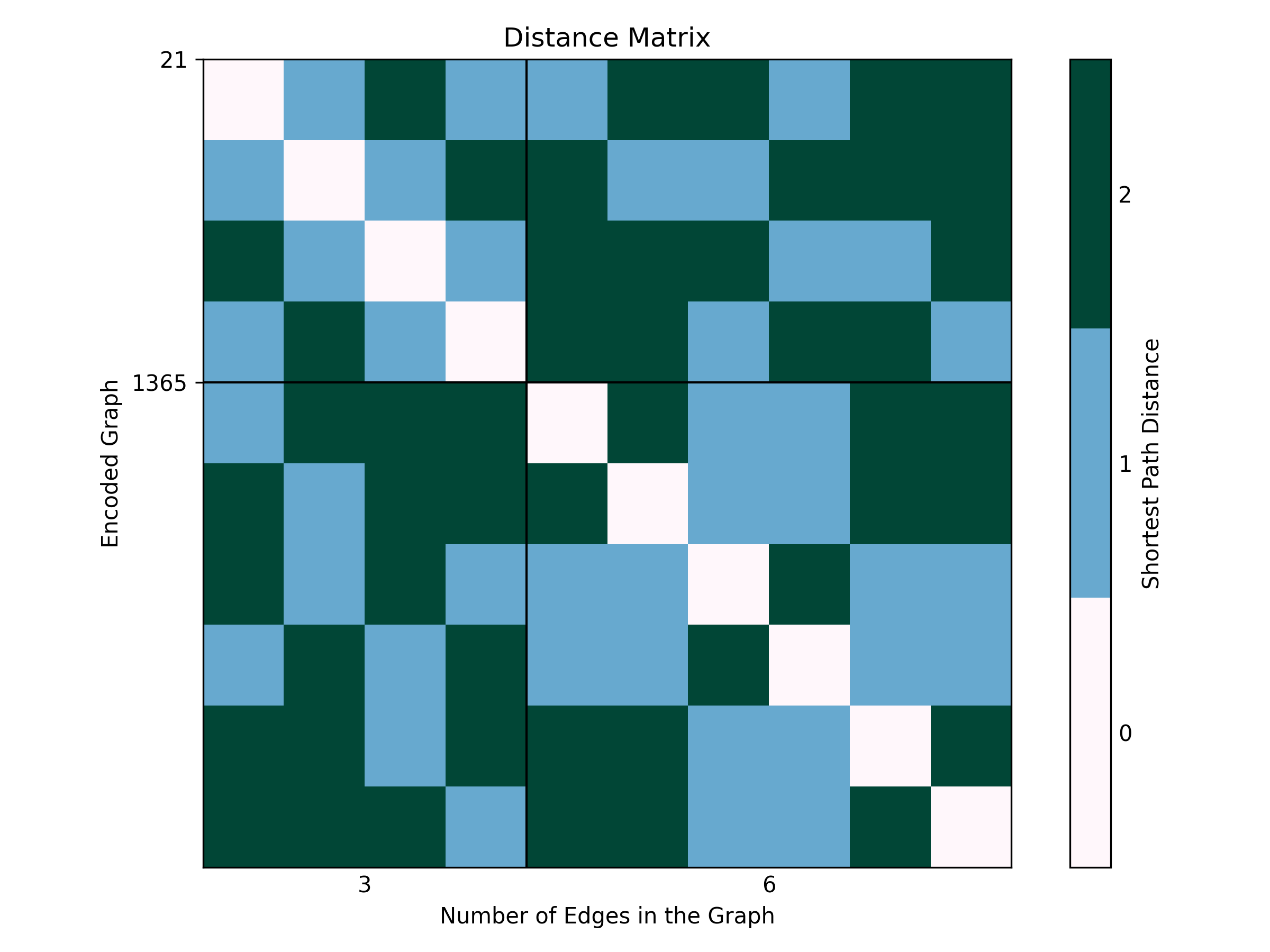}
        \subcaption{Orbit No. 2}
    \end{minipage}
    \hfill
    \begin{minipage}[t]{0.49\textwidth}
        \centering
        \includegraphics[width=\columnwidth]{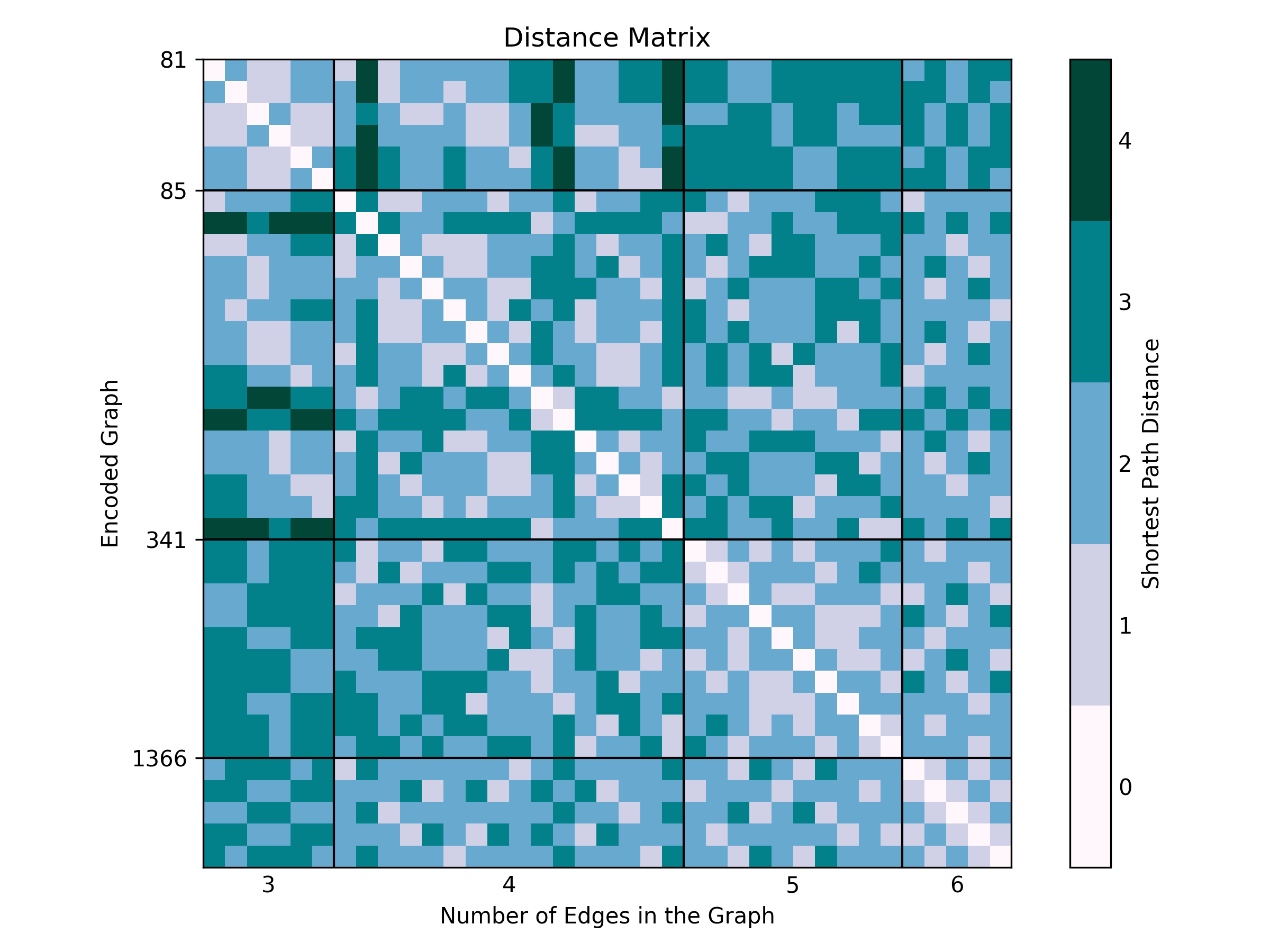}
        \subcaption{Orbit No. 3}
    \end{minipage}

    \caption{An example of the OG of orbit No. 2 (top) along with the distance matrices of orbit No. 2 (a) and orbit No. 3 (b).}
    \label{fig:composite}
\end{figure*}

In Figure \ref{fig:composite}, we give as an example the OG of orbit 2. Additionally, we give as an example the distance matrices for orbit 2 and orbit 3.

\subsection{Physical Interpretation of the observables and their statistics}
Bearing in mind the previously presented details, we present the observables for each one of the entanglement classes for $n\leq 7$ for qutrits in Table \ref{tab:summary_statistics_1}. Apart from the Schmidt measure $E_S$, which is an entanglement monotone tailored to describe graph states \cite{Eisert}, the physical implications of the used observables must be discussed. It must be underlined that the ordering of the orbits has been done first with respect to the number of particles each entanglement class has. Next, for each orbit we find the graph with firstly the minimum number of edges, in case of a tie we take the graph with the smaller total weight, if we still have a tie we take the smallest encoding. For each orbit we call this its representative $g(v,e)$, and create a tuple $(|e|,\text{total\_weight}(g),\text{encoding}(g))$ and sort the orbits based on this tuple. 

\begin{table*}[t!]
\centering
\caption{A presentation of the observables for each entanglement class for $n\leq 7$ qutrits. By orbit, we denote the adopted indexing of each one of the orbits. The column $|V|$ corresponds to the number of vertices $|V|$ of the OG, $|e|$ corresponds to the number of edges the representative of the entanglement class has. The Schmidt measure of each entanglement class is denoted as $E_S$, in cases where an exact value was not obtained, lower and upper bounds are given.
We adopt a compact notation for the lower bound $a$ and upper bound $b$ as $(a,b)$. For the number of loops $N_{\mathcal{L}}$, we rescaled the data with the natural logarithm plus one. $\chi_{OG}$ is the chromatic number of OG, and $\chi_i$ is the minimum chromatic number of a state in an orbit. $\mathcal{D}$ is the density of OG, $\langle d_{OG} \rangle$ is the ASPL of OG, $d_{OG}^{\text{max}} $ is the diameter of OG, $\text{deg}(OG)_{\text{max}}$ is the maximum degree of an OG node and finally, $\deg(g)_{\text{min}}$ which is the minimal maximum degree of one node of a state in the orbit.}
\label{tab:summary_statistics_1}
\resizebox{\textwidth}{!}{
\begin{tabular}{ccccccccccccc}
\toprule
orbit &  $n$ &$|V|$& $|e|$ &$\chi_{OG}$ & $\text{ln}\left(N_{\mathcal{L}}+1\right)$ & $\chi_i$ & $\mathcal{D}$ & $\langle d_{OG} \rangle$ &  $d_{OG}^{\text{max}} $ & $\deg(g)_{\text{min}}$ & $\text{deg}(OG)_{\text{max}}$ & $E_S$ \\
\midrule
1 & $3$ & $7$ & $2$ & $3$ & $1.79$ & $2$ & $0.90476$ & $1.33$ & $2$ & $2$ & $5$ & $1$ \\
2 & $4$ & $10$ & $3$ & $3$ & $1.61$ & $2$ & $0.53333$ & $1.56$ & $2$ & $3$ & $6$ & $1$ \\
3 & $4$ & $37$ & $3$ & $4$ & $2.94$ & $2$ & $0.21021$ & $2.18$ & $4$ & $3$ & $9$ & $2$ \\
4 & $4$ & $6$ & $4$ & $4$ & $1.95$ & $2$ & $1.13333$ & $1.27$ & $2$ & $5$ & $4$ & $2$ \\
5 & $5$ & $11$ & $4$ & $3$ & $2.08$ & $2$ & $0.58182$ & $1.58$ & $3$ & $4$ & $6$ & $1$ \\
6 & $5$ & $88$ & $4$ & $6$ & $3.99$ & $2$ & $0.10580$ & $2.55$ & $4$ & $4$ & $11$ & $2$ \\
7 & $5$ & $255$ & $4$ & $7$ & $5.08$ & $2$ & $0.04502$ & $3.01$ & $5$ & $4$ & $14$ & $2$ \\
8 & $5$ & $219$ & $5$ & $8$ & $3.83$ & $2$ & $0.06070$ & $2.66$ & $6$ & $3$ & $15$ & (2, 3) \\
9 & $5$ & $139$ & $5$ & $7$ & $4.26$ & $2$ & $0.08675$ & $2.53$ & $4$ & $6$ & $15$ & $2$ \\
10 & $6$ & $14$ & $5$ & $3$ & $1.95$ & $2$ & $0.40659$ & $1.81$ & $3$ & $5$ & $6$ & $1$ \\
11 & $6$ & $116$ & $5$ & $5$ & $4.11$ & $2$ & $0.07901$ & $2.76$ & $5$ & $5$ & $12$ & $2$ \\
12 & $6$ & $66$ & $5$ & $4$ & $3.61$ & $2$ & $0.13520$ & $2.46$ & $5$ & $4$ & $11$ & $2$ \\
13 & $6$ & $666$ & $5$ & $7$ & $5.94$ & $2$ & $0.01830$ & $3.52$ & $6$ & $4$ & $14$ & $2$ \\
14 & $6$ & $420$ & $5$ & $7$ & $5.59$ & $2$ & $0.02875$ & $3.28$ & $5$ & $3$ & $15$ & $3$ \\
15 & $6$ & $1850$ & $5$ & $8$ & $6.90$ & $2$ & $0.00787$ & $3.85$ & $7$ & $3$ & $17$ & $3$ \\
16 & $6$ & $162$ & $6$ & $5$ & $4.69$ & $2$ & $0.06970$ & $2.79$ & $5$ & $4$ & $15$ & $2$ \\
17 & $6$ & $164$ & $6$ & $6$ & $4.69$ & $2$ & $0.06793$ & $2.84$ & $5$ & $4$ & $15$ & $3$ \\
18 & $6$ & $4698$ & $6$ & $9$ & $7.26$ & $2$ & $0.00339$ & $4.16$ & $7$ & $4$ & $18$ & $3$ \\
19 & $6$ & $7572$ & $6$ & $10$ & $0.00$ & $2$ & $0.00229$ & $4.17$ & $9$ & $4$ & $18$ & $3$ \\
20 & $6$ & $180$ & $6$ & $8$ & $4.78$ & $2$ & $0.07337$ & $2.61$ & $5$ & $6$ & $15$ & $2$ \\
21 & $6$ & $331$ & $6$ & $8$ & $5.28$ & $2$ & $0.04103$ & $2.90$ & $5$ & $5$ & $17$ & $2$ \\
22 & $6$ & $164$ & $6$ & $6$ & $4.69$ & $2$ & $0.06793$ & $2.84$ & $5$ & $4$ & $15$ & $3$ \\
23 & $6$ & $1008$ & $6$ & $8$ & $6.24$ & $2$ & $0.01393$ & $3.50$ & $6$ & $4$ & $17$ & $3$ \\
24 & $6$ & $980$ & $6$ & $10$ & $6.39$ & $2$ & $0.01635$ & $3.34$ & $6$ & $4$ & $18$ & $3$ \\
25 & $6$ & $1892$ & $6$ & $10$ & $5.46$ & $2$ & $0.00915$ & $3.42$ & $6$ & $4$ & $18$ & $3$ \\
26 & $6$ & $1200$ & $7$ & $10$ & $5.24$ & $2$ & $0.01460$ & $3.15$ & $5$ & $5$ & $18$ & $3$ \\
27 & $6$ & $1572$ & $7$ & $10$ & $6.24$ & $2$ & $0.01012$ & $3.51$ & $6$ & $4$ & $18$ & $3$ \\
28 & $6$ & $119$ & $8$ & $7$ & $4.01$ & $2$ & $0.12648$ & $2.21$ & $4$ & $6$ & $18$ & $3$ \\
29 & $6$ & $462$ & $9$ & $9$ & $0.00$ & $2$ & $0.03276$ & $2.84$ & $6$ & $4$ & $18$ & (3, 4) \\
30 & $6$ & $940$ & $9$ & $9$ & $3.22$ & $2$ & $0.01706$ & $3.18$ & $5$ & $5$ & $18$ & $3$ \\
31 & $7$ & $15$ & $6$ & $3$ & $2.30$ & $2$ & $0.42857$ & $1.87$ & $4$ & $6$ & $6$ & $1$ \\
32 & $7$ & $137$ & $6$ & $6$ & $4.41$ & $2$ & $0.07042$ & $2.86$ & $5$ & $5$ & $12$ & $2$ \\
33 & $7$ & $156$ & $6$ & $8$ & $4.54$ & $2$ & $0.06352$ & $2.90$ & $5$ & $4$ & $12$ & $2$ \\
34 & $7$ & $852$ & $6$ & $7$ & $6.25$ & $2$ & $0.01473$ & $3.66$ & $6$ & $4$ & $15$ & $2$ \\
35 & $7$ & $483$ & $6$ & $7$ & $5.68$ & $2$ & $0.02554$ & $3.36$ & $6$ & $3$ & $15$ & $2$ \\
36 & $7$ & $565$ & $6$ & $8$ & $5.97$ & $2$ & $0.02261$ & $3.40$ & $6$ & $4$ & $16$ & $3$ \\
37 & $7$ & $1069$ & $6$ & $8$ & $6.59$ & $2$ & $0.01242$ & $3.71$ & $6$ & $3$ & $17$ & $3$ \\
38 & $7$ & $5052$ & $6$ & $9$ & $7.94$ & $2$ & $0.00302$ & $4.42$ & $7$ & $3$ & $18$ & $3$ \\
39 & $7$ & $5853$ & $6$ & $9$ & $8.27$ & $2$ & $0.00271$ & $4.46$ & $7$ & $3$ & $20$ & $3$ \\
40 & $7$ & $1155$ & $6$ & $8$ & $6.73$ & $2$ & $0.01270$ & $3.71$ & $7$ & $3$ & $20$ & $3$ \\
41 & $7$ & $13923$ & $6$ & $10$ & $8.94$ & $2$ & $0.00127$ & $4.72$ & $8$ & $2$ & $21$ & $3$ \\
42 & $7$ & $576$ & $7$ & $8$ & $6.04$ & $2$ & $0.02250$ & $3.39$ & $6$ & $4$ & $17$ & $2$ \\
43 & $7$ & $577$ & $7$ & $8$ & $6.05$ & $3$ & $0.02253$ & $3.39$ & $6$ & $4$ & $17$ & $3$ \\
44 & $7$ & $6451$ & $7$ & $9$ & $7.70$ & $3$ & $0.00255$ & $4.29$ & $7$ & $4$ & $18$ & $3$ \\
45 & $7$ & $3144$ & $7$ & $9$ & $7.71$ & $2$ & $0.00495$ & $4.15$ & $7$ & $3$ & $20$ & $3$ \\

\end{tabular}
}
\end{table*}

\begin{table*}
\vspace{-1cm}
\centering
\label{tab:summary_statistics_2}
\resizebox{\textwidth}{!}{
\begin{tabular}{ccccccccccccc}
\toprule
orbit &  $n$ &$|V|$& $|e|$ &$\chi_{OG}$ & $\text{ln}\left(N_{\mathcal{L}}+1\right)$ & $\chi_i$ & $\mathcal{D}$ & $\langle d_{OG} \rangle$ &  $d_{OG}^{\text{max}}$ & $\deg(g)_{\text{min}}$ & $\text{deg}(OG)_{\text{max}}$ & $E_S$ \\
\midrule

46 & $7$ & $14703$ & $7$ & $10$ & $8.95$ & $3$ & $0.00120$ & $4.68$ & $7$ & $3$ & $21$ & $3$ \\
47 & $7$ & $3153$ & $7$ & $9$ & $7.70$ & $3$ & $0.00493$ & $4.15$ & $7$ & $3$ & $20$ & $3$ \\
48 & $7$ & $14724$ & $7$ & $10$ & $8.94$ & $3$ & $0.00120$ & $4.68$ & $7$ & $3$ & $21$ & $3$ \\
49 & $7$ & $71046$ & $7$ & $11$ & $9.96$ & $2$ & $0.00027$ & $5.34$ & $8$ & $3$ & $21$ & $3$ \\
50 & $7$ & $35091$ & $7$ & $11$ & $9.37$ & $3$ & $0.00055$ & $5.01$ & $8$ & $3$ & $21$ & (3, 4) \\
51 & $7$ & $48207$ & $7$ & $12$ & $6.18$ & $3$ & $0.00043$ & $4.78$ & $8$ & $2$ & $21$ & (3, 4) \\
52 & $7$ & $239$ & $7$ & $7$ & $4.88$ & $2$ & $0.05401$ & $2.82$ & $5$ & $5$ & $15$ & $2$ \\
53 & $7$ & $896$ & $7$ & $9$ & $6.23$ & $2$ & $0.01584$ & $3.41$ & $6$ & $4$ & $18$ & $2$ \\
54 & $7$ & $577$ & $7$ & $7$ & $6.05$ & $3$ & $0.02253$ & $3.39$ & $6$ & $4$ & $17$ & $3$ \\
55 & $7$ & $2695$ & $7$ & $9$ & $7.31$ & $2$ & $0.00553$ & $4.03$ & $6$ & $4$ & $18$ & $3$ \\
56 & $7$ & $1597$ & $7$ & $9$ & $6.93$ & $2$ & $0.01041$ & $3.61$ & $6$ & $4$ & $21$ & $3$ \\
57 & $7$ & $3132$ & $7$ & $8$ & $7.71$ & $2$ & $0.00502$ & $4.07$ & $6$ & $4$ & $20$ & $3$ \\
58 & $7$ & $4872$ & $7$ & $10$ & $7.87$ & $2$ & $0.00349$ & $4.12$ & $7$ & $4$ & $21$ & $3$ \\
59 & $7$ & $1360$ & $7$ & $10$ & $6.63$ & $2$ & $0.01194$ & $3.52$ & $6$ & $3$ & $18$ & $3$ \\
60 & $7$ & $3153$ & $7$ & $10$ & $7.70$ & $3$ & $0.00493$ & $4.15$ & $7$ & $3$ & $20$ & $3$ \\
61 & $7$ & $14724$ & $7$ & $10$ & $8.94$ & $3$ & $0.00120$ & $4.68$ & $7$ & $3$ & $21$ & (3, 4) \\
62 & $7$ & $14679$ & $7$ & $10$ & $8.99$ & $2$ & $0.00120$ & $4.75$ & $7$ & $3$ & $21$ & $3$ \\
63 & $7$ & $35418$ & $7$ & $11$ & $9.32$ & $2$ & $0.00054$ & $4.96$ & $8$ & $3$ & $21$ & $3$ \\
64 & $7$ & $7440$ & $7$ & $11$ & $8.30$ & $2$ & $0.00253$ & $4.28$ & $7$ & $3$ & $21$ & $3$ \\
65 & $7$ & $48207$ & $7$ & $12$ & $6.18$ & $3$ & $0.00043$ & $4.78$ & $9$ & $3$ & $21$ & (3, 4) \\
66 & $7$ & $71244$ & $8$ & $11$ & $10.00$ & $2$ & $0.00027$ & $5.33$ & $8$ & $3$ & $21$ & $3$ \\
67 & $7$ & $57072$ & $8$ & $12$ & $6.20$ & $3$ & $0.00036$ & $4.90$ & $8$ & $3$ & $21$ & (3, 4) \\
68 & $7$ & $71400$ & $8$ & $11$ & $10.03$ & $3$ & $0.00027$ & $5.33$ & $8$ & $3$ & $21$ & (3, 4) \\
69 & $7$ & $357$ & $8$ & $9$ & $5.61$ & $2$ & $0.04068$ & $2.91$ & $5$ & $5$ & $19$ & $2$ \\
70 & $7$ & $7818$ & $8$ & $11$ & $7.80$ & $3$ & $0.00239$ & $4.10$ & $6$ & $4$ & $21$ & $3$ \\
71 & $7$ & $2162$ & $8$ & $10$ & $6.68$ & $3$ & $0.00762$ & $3.66$ & $6$ & $4$ & $18$ & $3$ \\
72 & $7$ & $2062$ & $8$ & $9$ & $7.24$ & $3$ & $0.00770$ & $3.74$ & $6$ & $4$ & $20$ & $3$ \\
73 & $7$ & $29340$ & $8$ & $10$ & $9.64$ & $3$ & $0.00061$ & $5.08$ & $7$ & $3$ & $21$ & $3$ \\
74 & $7$ & $15600$ & $8$ & $11$ & $8.48$ & $3$ & $0.00121$ & $4.43$ & $7$ & $4$ & $21$ & $3$ \\
75 & $7$ & $37260$ & $8$ & $12$ & $7.16$ & $2$ & $0.00055$ & $4.70$ & $7$ & $4$ & $21$ & $3$ \\
76 & $7$ & $2631$ & $8$ & $9$ & $7.47$ & $2$ & $0.00639$ & $3.94$ & $7$ & $4$ & $21$ & $3$ \\
77 & $7$ & $23634$ & $8$ & $11$ & $8.99$ & $3$ & $0.00082$ & $4.70$ & $7$ & $4$ & $21$ & (3, 4) \\
78 & $7$ & $35589$ & $8$ & $11$ & $9.38$ & $3$ & $0.00054$ & $4.96$ & $8$ & $3$ & $21$ & (3, 4) \\
79 & $7$ & $11802$ & $8$ & $11$ & $8.31$ & $3$ & $0.00162$ & $4.38$ & $7$ & $3$ & $21$ & (3, 4) \\
80 & $7$ & $71400$ & $8$ & $11$ & $10.03$ & $3$ & $0.00027$ & $5.33$ & $8$ & $3$ & $21$ & (3, 4) \\
81 & $7$ & $338868$ & $8$ & $12$ & $6.15$ & $3$ & $0.00006$ & $5.89$ & $9$ & $3$ & $21$ & (3, 4) \\
82 & $7$ & $1030$ & $8$ & $10$ & $6.49$ & $2$ & $0.01906$ & $3.13$ & $6$ & $4$ & $21$ & $3$ \\
83 & $7$ & $10084$ & $9$ & $11$ & $5.83$ & $2$ & $0.00200$ & $4.08$ & $8$ & $3$ & $21$ & $3$ \\
84 & $7$ & $59454$ & $9$ & $11$ & $5.77$ & $3$ & $0.00034$ & $4.99$ & $9$ & $3$ & $21$ & (3, 4) \\
85 & $7$ & $19872$ & $9$ & $12$ & $5.75$ & $2$ & $0.00102$ & $4.42$ & $7$ & $4$ & $21$ & $3$ \\
86 & $7$ & $5016$ & $9$ & $10$ & $7.92$ & $3$ & $0.00349$ & $4.08$ & $7$ & $4$ & $21$ & $3$ \\
87 & $7$ & $59454$ & $9$ & $12$ & $5.77$ & $3$ & $0.00034$ & $4.99$ & $9$ & $4$ & $21$ & (3, 4) \\
88 & $7$ & $170580$ & $9$ & $12$ & $7.79$ & $3$ & $0.00012$ & $5.50$ & $9$ & $4$ & $21$ & (3, 4) \\
89 & $7$ & $173784$ & $9$ & $12$ & $6.18$ & $3$ & $0.00012$ & $5.52$ & $8$ & $4$ & $21$ & (3, 4) \\
90 & $7$ & $9669$ & $9$ & $12$ & $7.97$ & $3$ & $0.00213$ & $4.11$ & $7$ & $4$ & $21$ & (3, 4) \\
91 & $7$ & $4107$ & $9$ & $10$ & $7.64$ & $2$ & $0.00458$ & $3.87$ & $7$ & $4$ & $21$ & $3$ \\
92 & $7$ & $2136$ & $9$ & $11$ & $5.78$ & $3$ & $0.00912$ & $3.38$ & $6$ & $4$ & $21$ & $3$ \\
93 & $7$ & $15288$ & $10$ & $10$ & $8.50$ & $3$ & $0.00123$ & $4.52$ & $8$ & $4$ & $21$ & (3, 4) \\
94 & $7$ & $88380$ & $10$ & $12$ & $7.14$ & $3$ & $0.00023$ & $5.18$ & $8$ & $3$ & $21$ & (3, 4) \\
95 & $7$ & $58908$ & $10$ & $12$ & $6.20$ & $3$ & $0.00035$ & $4.88$ & $8$ & $3$ & $21$ & (3, 4) \\
96 & $7$ & $8214$ & $10$ & $11$ & $7.87$ & $3$ & $0.00231$ & $4.12$ & $7$ & $4$ & $21$ & $3$ \\
97 & $7$ & $24702$ & $10$ & $11$ & $8.98$ & $3$ & $0.00077$ & $4.74$ & $8$ & $4$ & $21$ & (3, 4) \\
98 & $7$ & $114528$ & $10$ & $12$ & $6.96$ & $3$ & $0.00018$ & $5.29$ & $8$ & $4$ & $21$ & (3, 4) \\
99 & $7$ & $175464$ & $10$ & $12$ & $6.15$ & $3$ & $0.00012$ & $5.52$ & $9$ & $4$ & $21$ & (3, 4) \\
100 & $7$ & $58908$ & $10$ & $12$ & $6.20$ & $3$ & $0.00035$ & $4.90$ & $8$ & $4$ & $21$ & (3, 4) \\
101 & $7$ & $38388$ & $10$ & $12$ & $5.13$ & $3$ & $0.00054$ & $4.59$ & $7$ & $4$ & $21$ & (3, 4) \\
102 & $7$ & $16745$ & $11$ & $12$ & $5.24$ & $3$ & $0.00124$ & $4.16$ & $7$ & $4$ & $21$ & (3, 4) \\
103 & $7$ & $16745$ & $11$ & $12$ & $5.24$ & $3$ & $0.00124$ & $4.16$ & $7$ & $4$ & $21$ & (3, 4) \\
\bottomrule
\end{tabular}
}
\end{table*}
 
A key idea to grasp is that every LC operation for qutrits can be represented as a sequence of local complementations and local scalings. So, the following discussion assumes that the only local operations we can apply are these two. Taking this into account, one can easily understand that the number of self-loops for the OG reflects the number of states in an entanglement class where at least one of the two operations in at least one of the particles will lead to the same state. The minimum chromatic number of a graph state in an entanglement class is directly connected to the Schmidt Measure $E_S$ as indicated in \cite{Eisert} for qubits and in \cite{our_col_paper} for qudits. The maximum degree of the OG demonstrates the maximum number of states one can obtain using only a single local operation. As we have discussed in Appendix \ref{app:quant-def}, the maximum degree of a multi-graph and the chromatic index are tightly connected. As examined in \cite{opt_grap_state_prep}, determining the minimum edge color of a state in an entanglement class corresponds to the minimum number of steps required to prepare it by using $CZ$ operations for qubits. A similar philosophy can be applied to qutrits as long as we are only allowed to assume $CZ$ operations and not $CZ^2$. This means that the more the local dimension increases, the less the chromatic index depicts the depth of the preparation circuit of a graph state. For this reason, it crucial to determine the minimum across the maximum degrees of each state in an entanglement class denoted as $\deg(g)_{\text{min}}$. 

We proceed with our discussion with the implication of the density of the OG, denoted as $\mathcal{D}$. Considering the definition of $\mathcal{D}$, it is straightforward to understand that the closer it is to one, the more the chances of transforming from one state to another using only one of the local operators increase. At this juncture, we turn our focus to the ASPL of the OG denoted as $\langle d_{OG} \rangle$. It reveals the average number of local operations one has to apply to transform one state to another in the orbit. This observable, combined with the minimum across the maximum degrees of each state in an entanglement class, is an indicator of the preparation complexity of every state in a specific entanglement class. The diameter of the OG which mathematically is connected with the distance matrix as $\max\limits_{\forall i,j \in E} \{ d_{OG}^{ij} \}$, frequently denoted as $d_{OG}^{\text{max}}$, represents the maximum number of local operations one will need to transform from a state to any other one in the same orbit. Last but not least is the chromatic number of the OG, which reflects the number of different sets of graph states that are not possible to transform with a single local operation.

Bearing the above information in mind, some useful statistical properties of the observables are presented in Table \ref{tab:summary_statistics_obs}. These statistical properties will assist us in our ultimate endeavor in this work, to find physical implications for the entanglement classes of qutrits via their orbit exploration. The adopted order in this case is crucial since it can impact some of the statistical properties.

Let us start by discussing the chromatic number of the OG, $\chi_{OG}$. This distribution is slightly left-skewed, a fact justified by the median and mode both being slightly higher than the average, as well as the value of skewness itself. The values are relatively symmetrically distributed around $10$ with a moderate spread. These facts indicate that the number of different sets that are not connected through one local operation is, in general, relatively large for all entanglement classes. The next observable to discuss is the number of self-loops. Our findings here suggest that the distribution has heavy right tails with significant outliers and a very wide spread, pulling the mean far from the median. This indicates that this property is not connected with the properties of the entanglement class, a result that will become apparent when the correlation coefficients are presented. 

We continue our analysis with statistical properties of the dataset describing the minimum chromatic number in each entanglement class. In this case, as expected, the distribution is flat and packed since the two possible values for the orbit were either $2$ or $3$. We also observe that a weak majority of orbits have $\chi_i=2$. The next one is the density of the OG $\mathcal{D}$, where it becomes clear that we have graphs with, in general, very low density, considering the mean, the median, and the mode of the dataset. Even in this case, we observe that the orbit $4$, which is the GHZ orbit \clearpage \noindent for $n=4$, has $\mathcal{D}=1.13333$. This happens because of the existence of self-loops. A general observation stemming from $\mathcal{D}$ is that OG are typically sparse graphs, indicating that transforming from one state to another with a single local operation in the same entanglement class is rare. In general, there is a strong right skew and heavy tails, and as we proceed, the density of the graph decreases logarithmically.
\begin{table*}
\centering
\caption{Summary of the statistical properties of the considered observables including the mean, standard deviation, variance, median, mode, range, the difference between the 75th percentile and 25th percentile called Interquartile Range (IQR), skewness, and kurtosis.}
\label{tab:summary_statistics_obs}
\resizebox{\textwidth}{!}{
\begin{tabular}{cccccccccc}
\toprule
 & Mean & Standard Deviation & Variance & Median & Mode & Range & IQR & Skewness & Kurtosis \\
\midrule
$\chi_{OG}$& $9.12$ & $2.45$ & $6.01$ & $10.00$ & $10.00$ & $9.00$ & $3.00$ & $-0.89$ & $0.14$ \\
$N_{\mathcal{L}}$ & $2674.96$ & $4912.78$ & $24135405.61$ & $507.00$ & $108.00$ & $22632.00$ & $2245.00$ & $2.77$ & $7.46$ \\
$\chi_i$ & $2.40$ & $0.49$ & $0.24$ & $2.00$ & $2.00$ & $1.00$ & $1.00$ & $0.42$ & $-1.83$ \\
$\mathcal{D}$  & $0.0571$ & $0.1675$ & $0.0281$ & $0.0050$ & $0.0003$ & $1.1333$ & $0.0230$ & $4.5329$ & $21.8993$ \\
$\langle d_{OG} \rangle$ & $3.85$ & $1.03$ & $1.05$ & $4.07$ & $2.84$ & $4.62$ & $1.53$ & $-0.40$ & $-0.37$ \\
$d_{OG}^{\text{max}}$  & $6.42$ & $1.55$ & $2.40$ & $7.00$ & $7.00$ & $7.00$ & $1.00$ & $-0.67$ & $0.50$ \\
$\deg(g)_{\text{min}}$ & $3.77$ & $0.80$ & $0.64$ & $4.00$ & $4.00$ & $4.00$ & $1.00$ & $0.56$ & $0.78$ \\
$\text{deg}(OG)_{\text{max}}$ & $17.96$ & $4.21$ & $17.69$ & $20.00$ & $21.00$ & $17.00$ & $4.00$ & $-1.67$ & $2.23$ \\
$E_S$ & $2.84$ & $0.65$ & $0.42$ & $3.00$ & $3.00$ & $2.50$ & $0.50$ & $-1.24$ & $0.95$ \\
\bottomrule
\end{tabular}
}
\end{table*}

We turn our focus on the $\langle d_{OG} \rangle$. We find that it has a moderate spread and is mildly left-skewed. Additionally, as we increase the orbit numbering, the $\langle d_{OG} \rangle$ increases. Even if $\langle d_{OG} \rangle$ is a float, the number of local operations is an integer, and these results suggest that in general, after applying $3$ or $4$ we should be able to transform one state to another. However, this is not true in any case, and the diameter of the OG, $d_{OG}^{\text{max}}$, must be studied. We realize that, on average, the diameter is either 8 or 9, and the distribution is fairly symmetric with a tail to the left. A final remark is that our findings agree with \cite{Adcock2020mappinggraphstate}, where they found that the diameter for the qubit orbits for $n\leq 9$ was also at most $9$. From the current analysis, it is not clear if this is a coincidence or not, but it is something that could be examined in future research. 

We proceed with our work by discussing the minimum number of maximum degree of a node among the states of a specific entanglement class, $\deg(g)_{\text{min}}$. This observable is an indicator of the preparation complexity of the graph states or the entanglement class in general. The distribution is slightly right-skewed with a light right tail. The next observable is the maximum degree of the OG, which indicates the number of states we can obtain only with one local operation. This distribution is, in general, pulled left by lower outliers, exhibiting strong left skew and a median of $20$. It is worth mentioning that the mode of the data is $21$, and we observe that a large number of orbits for $n = 7$ have $\text{deg}(OG)_{\text{max}} = 21$, which is expected for $d = 3$. The final observable to discuss is the Schmidt Measure $E_S$, which is left-skewed but mostly clustered around 3, with a tight spread, given the range of $2.5$ and IQR of $0.5$. This observable will be crucial in the next part of our analysis. That is because it is an entanglement monotone tailored for graph states. Therefore, finding strong correlations between $E_S$ and any other of the observables, physical implications can be found for the entanglement classes, bearing in mind the corresponding physical interpretations.

\subsection{Correlations of the observables and physical implications}
Up to this point, it is apparent that the statistical examination of observable distributions can lead to useful physical implications. It is therefore natural to examine the correlation of these observables. In Subsection \ref{subsection:correlation-coef} the Pearson and Kendall's $\tau$ correlation coefficients were described. Bearing this in mind, Table \ref{tab:correlations-strong} can be presented, where only pairs with $|r|\geq 0.8$ were considered. A slightly stricter constraint from the typical $|r|\geq 0.7$ was adopted, because plenty of the observables exhibit outliers and knots leading to relatively high $r$ values. One crucial comment is that in some cases, significant improvement for both $r$ and $\tau$ was observed if one of the data sets was logarithmically rescaled. This means that if $|r|> 0.8$, the observables are (anti)correlated but not linearly. Before discussing each one of the findings, we have to note the encouraging fact that our purely quantum observable, $E_S$ has a strong correlation with the majority of the rest of the observables, already pointing out that studying the graph properties of OG can have interesting physical implications.

\begin{table*}[t]
\centering
\caption{Pairs of observables with strong Pearson correlation ($|r| > 0.8$), and their corresponding Kendall’s $\tau$ coefficients.}
\label{tab:correlations-strong}
\begin{tabular}{ccc}
\toprule
Compared observables & Pearson $r$ & Kendall $\tau$ \\
\midrule
$E_S$ vs $\chi_{OG}$ & 0.86 & 0.74 \\
$E_S$ vs $\ln\mathcal{D}$ & -0.82 & -0.71 \\
$E_S$ vs $\langle d_{OG} \rangle$ & 0.81 & 0.67 \\
$E_S$ vs $d_{OG}^{\text{max}}$ & 0.81 & 0.72 \\
$E_S$ vs $\deg(OG)_{\max}$ & 0.87 & 0.72 \\
$\deg(OG)_{\max}$ vs $\chi_{OG}$ & 0.92 & 0.81 \\
$\ln\deg(OG)_{\max}$ vs $\mathcal{D}$ & -0.89 & -0.76 \\
$\deg(OG)_{\max}$ vs $\langle d_{OG} \rangle$ & 0.85 & 0.71 \\
$\deg(OG)_{\max}$ vs $d_{OG}^{\text{max}}$ & 0.85 & 0.74 \\
$\langle d_{OG} \rangle$ vs $d_{OG}^{\text{max}}$ & 0.94 & 0.84 \\
$\ln\mathcal{D}$ vs $d_{OG}^{\text{max}}$ & -0.94 & -0.84 \\
$\mathcal{D}$ vs $\langle d_{OG} \rangle$ & -0.99 & -0.90 \\
$\chi_{OG}$ vs $\ln d_{OG}^{\text{max}}$ & 0.85 & 0.73 \\
$\chi_{OG}$ vs $\ln \langle d_{OG} \rangle$ & 0.88 & 0.69 \\
$\chi_{OG}$ vs $\ln \mathcal{D}$ & -0.91 & -0.78 \\
\bottomrule
\end{tabular}
\end{table*}

We will start by discussing our results related to $E_S$. The first one is $\left\{r\left(E_S,\chi_{OG}\right),\tau\left(E_S,\chi_{OG}\right)\right\} = \{0.86,0.74\}$ that justifies a strong correlation between $E_S$ and the OG chromatic number. A similar result was found for the qubit case by \cite{Adcock2020mappinggraphstate}. We proceed with $\left\{r\left(E_S, \text{ln}\mathcal{D} \right),\tau\left(E_S, \text{ln}\mathcal{D} \right)\right\} = \{-0.82 , -0.71\}$. Both correlation parameters admit that the two observables are strongly anti-correlated. Therefore entanglement classes where states have limited power to create many states via local operations seem to have higher $E_S$. We advance with $\left\{r\left(E_S,\langle d_{OG} \rangle\right),\tau\left(E_S,\langle d_{OG} \rangle\right)\right\} = \{0.81,0.67\}$. This is the only case in which $r>0.8$ but $\tau<0.7$, indicating a moderate correlation between the average number of operations required to transform one state to another and the $E_S$. The diameter of OG and $E_S$ are strongly correlated with $\left\{r\left(E_S,d_{OG}^{\text{max}}\right),\tau\left(E_S,d_{OG}^{\text{max}}\right)\right\} = \{0.81,0.72\}$. This indicates that entanglement classes with states that require multiple local operations to produce others in the orbit have larger $E_S$, a fact supported also by the moderate correlation observed between $E_S$ and $\langle d_{OG} \rangle$. A final comment is that we know that upper and lower bounds on $E_S$ can be found based on the number of particles of a state \cite{Eisert}. As we are exploring orbits with an increasing number of particles, the diameter of OG will also increase, especially for sparse graphs,  justifying our findings. The last observable that strongly correlates with $E_S$ is the maximum degree of the OG $\text{deg}(OG)_{\text{max}}$. We find that $\left\{r\left(E_S, \text{deg}(OG)_{\text{max}}\right),\tau\left(E_S, \text{deg}(OG)_{\text{max}}\right)\right\} = \{0.87, 0.72 \}$. This result illustrates that entanglement classes that have at least one state that can produce many states with a single local operation have higher $E_S$. This result is in agreement with the qubit case observed in \cite{Adcock2020mappinggraphstate}. 

Even if the rest of the correlation combinations do not involve $E_S$, they reflect interesting physical properties of the orbits. We observe that for orbits that include at least one state that can produce many others, the number of distinct set of states that are not connected via a single local operation also increases since $\left\{r\left(\text{deg}(OG)_{\text{max}}, \chi_{OG}\right),\tau\left(\text{deg}(OG)_{\text{max}}, \chi_{OG}\right)\right\} = \{0.92, 0.81\}$. Interestingly, $\ln \text{deg}(OG)_{\text{max}}$ is strongly anticorrelated with $\mathcal{D}$ which is a measure of the number of states that are connected with a single local operation in an entanglement class since $\left\{r\left(\ln \text{deg}(OG)_{\text{max}}, \mathcal{D}\right),\tau\left(\ln\text{deg}(OG)_{\text{max}}, \mathcal{D}\right)\right\} = \{-0.89, -0.76\}$. This is not surprising since the anticorrelation between $E_S$ and $\text{ln}\mathcal{D}$ and the correlation between $E_S$ and $\text{deg}(OG)_{\text{max}}$ have been established. Therefore, it becomes apparent that complementary information is presented regarding the nature and characteristics of the orbits. Since the physical explanation of the observables is already established, we will proceed with discussing a selection of correlation results. 

Interestingly, the preparation complexity of the entanglement class in terms of $CZ$ operations reflected by $\deg(g)_{\text{min}}$ does not appear to be connected with any of the other observables, raising questions to be addressed in future research. As expected, the average number of local operations to transform a state to another is strongly anti-correlated with the $\mathcal{D}$ since $\left\{r\left(\text{ln}\mathcal{D}, d_{OG}^{\text{max}}\right),\tau\left(\text{ln}\mathcal{D}, d_{OG}^{\text{max}} \right)\right\} = \{-0.94,-0.84\}$. A general comment is that the rest of the observables are intuitive for large and sparse graphs in graph theory. A final remark is that in the vast majority of cases, the Pearson coefficient was much more optimistic in detecting strong (anti)correlations while the Kendall's $\tau$ was more conservative, justifying our choice of correlation observables.

An additional note is regarding the comparison of our findings and the qubit case presented in \cite{Adcock2020mappinggraphstate}. It is noteworthy that the (anti)correlated observables for the qubit case also exhibit the same behavior for the qutrit case. This is a clear indication that studying the entanglement classes under the scope of the orbits and their graph-theoretic properties unveils physical properties that can be systematically investigated. On the other hand, some properties found for the qubit case will not be affected by focusing on higher dimensions. The reason for this is that every weighted graph of multiplicity $d-1$ will encapsulate all the graphs with multiplicity $d-2$, and thus, the aforementioned agreement is expected. For this reason, the computationally challenging task of deriving the orbits for $d=5$ or $d=7$ is not a necessity, in contrast with $d=3$, where orbits for $n\leq 7$ were obtained, to gain better insights for the statistical analysis of some of the observables and their correlations. Furthermore, we identified $9$ orbit pairs in total that have the same number of edges and vertices. For the majority of them, they had identical values for their observables. Therefore, for some of them, we explicitly confirmed they are isomorphic. For instance, for $n=6$ the orbits $17$ and $22$ from table \ref{tab:summary_statistics_1} and for $n=7$ the pair of orbits $43$ and $54$ as well as $47$ and $60$ are isomorphic to each other. An interesting fact is that, having a look at the representatives of those orbits, they have the same geometric structure but differ by one or two edges that have different weights. However, not every property of the OG for qubits is extended to qutrits. In our case, not a single orbit is a tree, but this was the only case a difference was observed for different local dimensions. 

\section{Discussion and Conclusions}\label{sec:discussion}

We have presented a comprehensive classification of qutrit graph state orbits under local complementation and local scaling, along with their statistical analysis. Our code also provides the tools to calculate the orbits for qubits, even if this is something already known in the literature, it was a required step to benchmark our code with the findings in \cite{Adcock2020mappinggraphstate, qubits_reps12}. Furthermore, our code can also be used to extract the orbits for $d>3$, but due to the computational difficulty a very small number of vertices can only be reached. Our statistical analysis reveals strong correlations between graph-theoretic properties of orbit graphs (OGs) and the entanglement characteristics of their associated states. Additionally, our work underlines that the OG can reveal vital information that can impact various aspects of quantum technologies and communications. For the rest of this section, some of them will be discussed.

The work we presented has applications in various aspects of quantum computing and technologies. One of them is quantum error correction and fault tolerance. Even if the construction of schemes with qutrits has already been proposed in the literature \cite{quditgraphcodeconstruction1,quditgraphcodeconstruction2}, the majority of optimization schemes are concentrated on the qubit case. Similarly to the qubit case, the extension to any local dimension requires the complete characterization of the orbit. In \cite{ft_graph_codes}, the importance of local complementation on finding the optimal graph for measurement-based loss tolerance was presented. The key idea is that only a single representative per entanglement class needs to be examined, significantly reducing computational overhead. However, this is true for logical measurements on an arbitrary basis. If loss-tolerant Pauli measurements are required, deviations in the performance of locally equivalent states are observed. In \cite{Tiurev_Eisert}, the importance of being equipped with the LC entanglement classes was underlined, a need that this work addresses for $d\geq3$ and $d$ a prime number. Another note is that this argument is valid only for the LC and not the LU classes. This happens because any form of Clifford conjugations on any Clifford operation will not distort the commutativity and thus will form legitimate code stabilizers, a detail crucial for the optimization of graph codes as noted in \cite{ft_graph_codes}.

For the qubit case, local complementation can be used for a more efficient state preparation \cite{phy_Adcock2018}. More concretely, using single-qubit gates has higher fidelity and is faster than two-qubit operations. Therefore, having a characterization for the LC classes is a critical step. Our work provides a necessary step for the development of theoretical protocols and physical realizations for the higher local dimension regime. Furthermore, in certain architectures for quantum computations, for instance, in linear optical quantum computing \cite{linear_optical_qc}, probabilistic generation of percolated resource states is observed. The aforementioned states are generated randomly and thus some are more useful than the others in the context of loss-tolerance or pathfinding \cite{Morley_Short_2018}. It is possible to optimize these implementations if the orbits for a specific local dimension are found. The previous are concentrated in the qubit case, and our work aims to provide a subtle yet vital step for the extension to qudits.

A crucial question on the entanglement study of graph states was the Local Unitary (LU) and LC (LU-LC) conjecture, which was proven to be false since a counterexample was found in \cite{ji2008lulcconjecturefalse}. In contrast, \cite{Eisert} established that the LU-LC conjecture is true for up to $n\leq 8$ qubits. A recent work improved the lower bound up to $n\leq 19$ particle states \cite{claudetLU}, where the LU-LC conjecture holds. However, the same question can be addressed for the qutrit case, and thus having a catalog with all the LC equivalent classes becomes a key step towards this direction, mainly for qutrits with $n\leq 7$. Additionally, having a lookup table for LC orbits has been proven useful in the study of the foliage partition, which is an LC-invariant that can be calculated efficiently \cite{Burchardt2025foliagepartition}. In this work, the authors were able to provide their invariant for small qubit graphs, while now one can perform the same calculation for qutrits.

Local complementation appears to be a key ingredient in the context of quantum networks \cite{networks_rev}. In this scenario, a graph state is distributed among spatially distant parties, and thus, two-particle operations are not allowed. In such settings, classical communication is often restricted to neighboring parties, as defined by the connectivity of the distributed graph state.
For the qubit case, utilizing local complementation provided a clear advantage since the number of required measurements was decreased \cite{networks} concerning the methods used in typical quantum repeater schemes \cite{qrepeaters}. The same thought process can be extended to qudits, considering the code performance can scale with the local dimension \cite{PhysRevLett_113_230501}, and having a look-up table with all the equivalent classes can, once again, assist in the optimization of the proposed protocols. 

This work is accompanied by a GitHub repository \cite{github} that provides the code to obtain the orbits, perform the statistical analysis, and visualize both the orbits and the desired statistical properties. While our methods generalize naturally to qutrit systems with $n \geq 8$ and to other prime dimensions $d \geq 3$, these pose significant computational challenges.
This work aims to assist in the development of theoretical and experimental implementations supporting the further development of quantum computing and technologies for qudits.

\acknowledgements
We would like to thank Otfried G\"uhne, Dorian Rudolph, and Evangelos Varvelis for the fruitful discussions.

\bibliography{qudit_lc}

\clearpage
\appendix
\onecolumngrid
\clearpage
\section{Quantities Definition}\label{app:quant-def}
In this section, the relevant concepts and quantities from graph theory are going to be introduced. A \textit{simple undirected graph} $(V,E)$ is defined as a set of vertices $V$ for which $V\subset\mathbb{N}$ and edges $E$ for which $E\subset [V]^2$. In the case of \textit{multigraphs}, the definition is almost identical to the previous one, except that E is a multiset, and thus there may be more than one edge between a pair of vertices. For simple graphs, like OG, we have a graph that can have \textit{self-loops} (edges connecting a vertex with itself), but having multiple edges between the vertices is impossible. In contrast, for \textit{multigraphs} it is possible to have multiple edges between the vertices. This is the case for the graphs describing \textit{qutrit} graph states, and thus, care must be taken for describing their mathematical properties. In the following, relevant quantities from graph theory are going to be presented, as well as the nomenclature used in our work. If a quantity is discussed and the extension for multigraphs is not presented, it means that the latter is straightforward.

\begin{itemize}
    \item Two vertices are called \textit{adjacent} if they belong to the same edge. 
    \item A \textit{walk} is a sequence of vertices and edges in a graph, and a \textit{path} is a walk where a vertex is visited only once.
    \item A \textit{cycle} is a walk such that the first and the last vertices are the same, all the vertices in between are distinct, and no edge is repeated.
    \item If a graph has no circles, it is called \textit{tree}. For qubits, tree graph states cannot serve as universal resources for measurement-based quantum computation (MBQC) \cite{tree_non_univqc}.
    \item For a specific vertex $v$ of the graph $G$ the \textit{neighborhood} ($N_G(i)$) is defined as the set of vertices the vertex $i$ is adjacent.
    \item If there is a path to connect every vertex with another one, the graph is called \textit{connected}.
    \item The \textit{adjacency matrix} $\Gamma$ is a symmetric $n \times n$ matrix defined over a finite field $ \mathbb{F}_d$, where each nonzero element $\Gamma_{ij}=\theta_{ij} \in \mathbb{F}_d$ specifies the weight between vertices $i$ and $j$ \cite{Eisert}:
\begin{equation}
\Gamma_{ij}= 
\begin{cases} 
\theta_{ij} & \text{if } \{i,j\} \in E  \\
0 & \text{otherwise } 
\end{cases}\ .
\end{equation}
\item If a graph can be placed on a two-dimensional plane without any of its edges crossing, it is called \textit{planar graph}.
\item The \textit{chromatic number} of a graph is defined as the minimum number of colors one can use such that no adjacent vertices have the same color \cite{graph_coloring}. It is often referred to as \textit{colorability} for the corresponding graphs of the graph state or, in general, vertex coloring. The chromatic number of OG will be denoted as $\chi_{OG}$. Additionally, the minimum colorability of a specific entanglement class will be denoted as $\chi_i$. This property is directly connected with the preparation complexity of the graph state as well as with quantum combinatorial designs, as it was demonstrated in \cite{our_col_paper}.

\item The \textit{distance} of two vertices of a simple graph is the number of edges of the shortest path between them. Since this notion will be used only for the OG it is not necessary to define the extension for multigraphs. The physical intuition distance between nodes of the OG is the minimum number of local scaling or local complementation operations required to transform one state into another within the same entanglement class.

\item The \textit{distance matrix} for the OG $d_{OG}$ has elements $d_{OG}^{ij}$ where $\{i,j\}\in E$ and $d_{OG}^{ii} = 0$. Two observables are emerging from this quantity, utilized in this work. The first one is the average of all the elements of $d_{OG}$, denoted as  $\langle d_{OG} \rangle$. The latter one is the \textit{diameter} of the OG which is defined as $\max\limits_{\forall \{i,j\} \in E} \{ d_{OG}^{ij} \}$, and often written as $d_{OG}^{\text{max}}$. These properties reflect the average and maximum number of local scaling and local complementation operations required to transform one state to another for a given entanglement class.

\item The notion of self-loops is already defined at the start of the section. In this work it the number of self-loops of a graph will be denoted as $N_{\mathcal{L}}$. Since this number will be orders of magnitude larger in comparison to the rest of the used quantities,  $\text{ln}\left(N_{\mathcal{L}}+1\right)$ will be used instead. Their existence in the OG indicates the number of times the application of local scaling or local complementation operation leads to the same graph for a specific entanglement class. 

\item The \textit{density} of a simple undirected graph $G=(V,E)$ will be denoted as $\mathcal{D}$ and it is defined as:
\begin{equation}
    \label{eq:desity_def}
    \mathcal{D} = \frac{\lvert E \rvert }{\binom{\lvert V \rvert }{2}} = \frac{2\lvert E \rvert }{\lvert V \rvert (\lvert V \rvert -1)}
\end{equation}

\item The \textit{vertex degree}, denoted as $\text{deg}(v)$ for $v\in V$, is the number of edges adjacent to a specific vertex of the graph. The extension for multigraphs is that it is the sum of the edge weights for edges incident to that vertex.

\item The \textit{chromatic index}, often referred to as edge chromatic number, is the minimum number of colors needed such that two edges on the same vertex do not have the same color. In the case of multigraphs, one should be careful that for edges with weight more than one, each weight one edge must have a different color.

\end{itemize}

It is clear that even if there is a physical interpretation of the above quantities, they are purely classical. For our work, it is crucial to have an observable reflecting a quantum property, therefore, an entanglement measure can be used. For graph states, the \textit{Schmidt measure} \cite{sm-def} is the tailored one to characterize them. One should recall that it is possible for any state vector $\ket{\psi}\in \mathcal{H}^{(1)}\otimes\cdots\otimes\mathcal{H}^{(n)}$ of a compound quantum system with n components to be written as:
\begin{equation}
\label{eq:ketSM}
    \ket{\psi} = \sum_{i=1}^{R}c_{i}\ket{\psi_i^{(1)}}\otimes\cdots\otimes\ket{\psi_i^{(n)}},\quad\text{where} \quad c_i\in\mathbb{C} \quad\text{and} \quad \ket{\psi_i^{(i)}}\in \mathcal{H}^{(i)} \quad\text{for} \quad i=1,\cdots, R.
\end{equation}
The Schmidt measure $E_S$ of the state $\ket{\psi}$ is defined as:
\begin{equation}
    \label{eq:sm-def}
    E_S(\ket{\psi}) = \log_2(r),
\end{equation}
where $r$ is the minimal number $R$ of the terms in the sum presented in (\ref{eq:ketSM}), over every linear decomposition into product states.

\newpage
\section{Representative states for each entanglement class}

\begin{figure}[ht!]
    \centering
    \includegraphics[width=\linewidth]{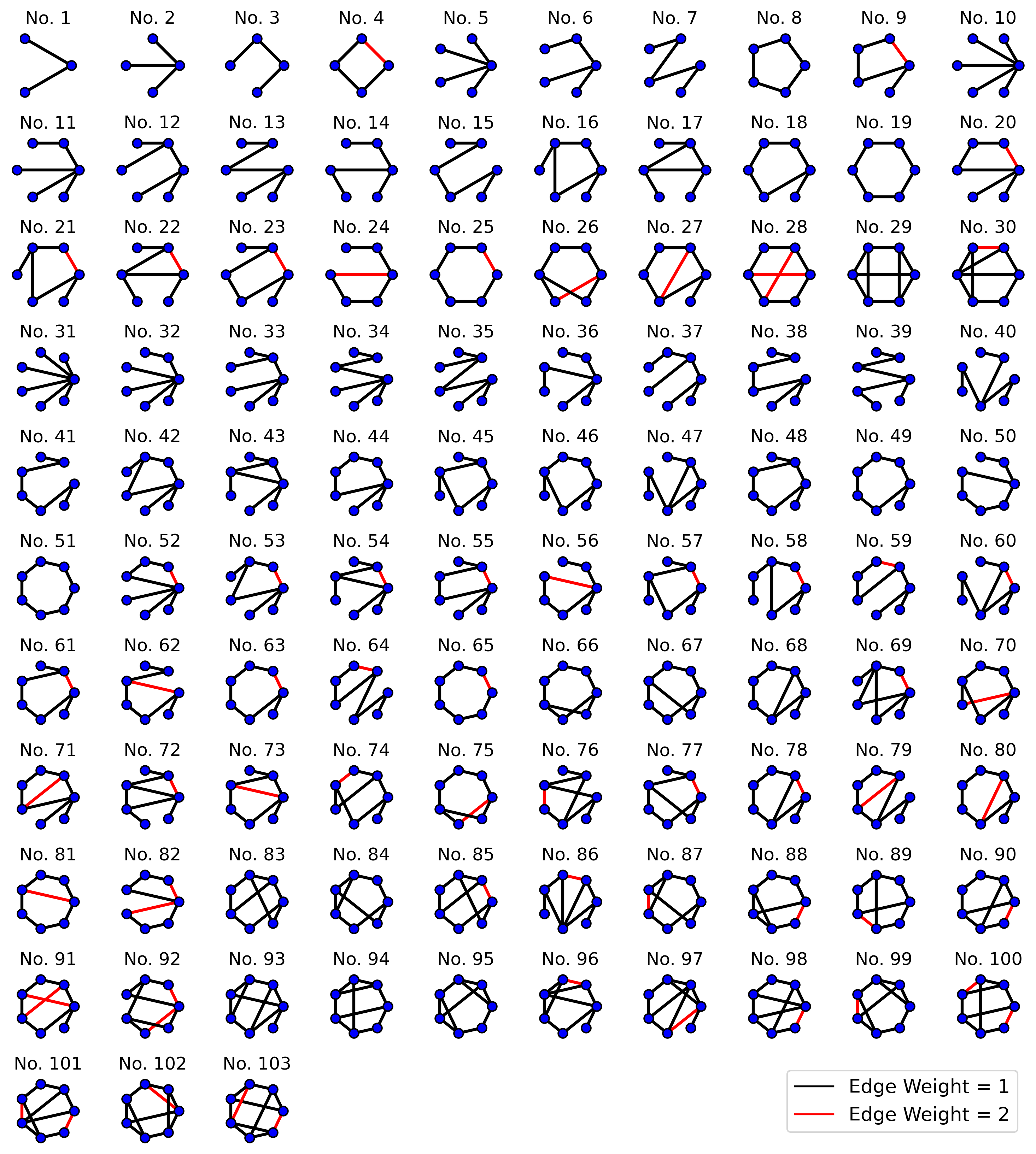}
    \caption{Representatives for qutrit graphs states up to $n=7$, sorted by number of particles, number of edges, total weight, and encoding value, in this specific order.}
    \label{fig:enter-label}
\end{figure}

\end{document}